\documentstyle[a4,12pt]{article}

\makeatletter
\@addtoreset{equation}{section}
\makeatother

\def\i{\ifmmode{\rm i}\else\char"10\fi}
\newcommand{\eps}{\epsilon}
\newcommand{\pdt}{\partial_t}
\newcommand{\pdx}{\partial_x}
\newcommand{\pdy}{\partial_y}
\newcommand{\pdz}{\partial_z}
\newcommand{\dxi}{\partial_\xi}
\newcommand{\deta}{\partial_\eta}
\newcommand{\dtau}{\partial_\tau}
\newcommand{\bu}{\mbox{\boldmath $u$}}
\newcommand{\bb}{\mbox{\boldmath $b$}}
\newcommand{\bok}{\mbox{\boldmath $k$}}
\newcommand{\br}{\mbox{\boldmath $r$}}
\newcommand{\bz}{\mbox{\boldmath$z$}}
\newcommand{\bB}{\mbox{\boldmath $B$}}
\newcommand{\bP}{\bar P}
\newcommand{\bc}{\bar c}
\newcommand{\baru}{\bar u}
\begin{document}
\title{Nonlinear Modulation of Travelling Rolls \\in Magnetoconvection}
\author{Ken-ichi Matsuba,Kenji Imai and Kazuhiro Nozaki\\
Department of Physics,Nagoya University,Nagoya 464-01,Japan}
\maketitle
\begin{abstract}
Modulational dynamics of oscillatory travelling rolls in magnetoconvection
is studied near the onset of a Hopf bifurcation. Using weakly nonlinear 
analysis, we derive an envelope equation of oscillatory travelling rolls in
the plane perpendicular to an ambient vertical magnetic field.The envelope 
equation is the Davey-Stewartson (DS) equation with complex coefficients, 
from which we obtain criteria for the modulational (Benjamin-Feir) 
instability of oscillatory travelling rolls. 
\end{abstract}
\section{Introduction}
A variety of spatially and temporally periodic patterns is found in 
weakly nonlinear Boussinesq convection in an imposed vertical magnetic 
field \cite{clune}. However, nonlinear modulation of the periodic patterns 
in the horizontal plane has thus far received less attention. Two kinds of
bifurcations are known to  convective patterns: one is to steady 
patterns and the other is to oscillatory patterns (a Hopf bifurcation) 
\cite{clune}.Near the onset of the former bifurcation, modulational 
dynamics of travelling rolls may be described by the Newell-Whitehead-Segel
(NWS) equation in terms of the Newell-Whitehead (NW) orderings
 \cite{newell1}. Near the onset of a Hopf bifurcation, the same ordings may 
 yield a more complicated envelope equation \cite{brand}, which includes 
 various terms of different orders.\\
 In this paper, introducing consistent orderings (different from the NW 
 orderings) to an envelope equation near the onset of a Hopf bifurcation, we 
 derive the DS equation with complex coefficients, where horizontal 
 incompressible flows couple to oscillatory travelling rolls in 
 magnetoconvection.The analysis of a instability of its spatially uniform 
 solution yields criteria for the modulational (Benjamin-Feir)
 instability of oscillatory travelling rolls.\par 
\section{Hopf Bifurcation}
Boussinesq convection in an imposed vertical magnetic field is described by
the equations \cite{clune}
\begin{eqnarray} 
\pdt\bu+(\bu\cdot\nabla)\bu&=&-\nabla P+\sigma R\Theta\bz+\sigma\zeta Q
(\nabla\times\bB)\times\bB+\sigma\triangle\bu,\label{b1}\\
\pdt\Theta+(\bu\cdot\nabla)\Theta&=&w+\triangle\Theta,\label{b2}\\
\pdt\bB+(\bu\cdot\nabla)\bB&=&(\bB\cdot\nabla)
\bu+\zeta\triangle\bB,\label{b3}
\end{eqnarray} 
together with
\begin{equation}
\nabla\cdot\bu=\nabla\cdot\bB=0.\label{b4}
\end{equation}
Here $\bu\equiv(u,v,w)$ is the dimensionless velocity in (x,y,z) 
coordinates, $\bz$ is the unit vector in the vertical ($z$)  
direction, $\Theta$ is the dimensionless temperature deviation from the 
conduction state and $\bB$ is the dimensionless magnetic field.
The parameters are the Prandtl number $\sigma$, the Rayleigh number $R$, 
the Chandrasekhar number $Q$ and the ratio of ohmic to thermal diffusivity
 denoted by $\zeta$. The magnetic field is given by  
$$\bB=\bz+\bb,$$
where $\bb\equiv(a,b,c)$. The boundary conditions are
$$\pdz u=\pdz v=w=\Theta=a=b=0\qquad \mbox{at}\qquad z=0,1.$$
The linear stability  analysis of the conduction state $\bu=\Theta=\bb=0$
shows that oscillatory convection sets in, for $\zeta<1$, at
$$R=R_0=\frac{(\pi^2+k^2)^3}{k^2}\frac{(\sigma+\zeta)(1+\zeta)}{\sigma}+
\frac{\pi^2+k^2}{k^2}\frac{\zeta(\sigma+\zeta)}{1+\sigma}\pi^2 Q.$$
Here k is the horizontal wavenumber determined by minimizing the critical 
Rayleigh number $R_0$. Thus $k=k_0$, where
$$(\pi^2+k_0^2)^3-\frac{3}{2}\pi^2(\pi^2+k_0^2)^2=\frac{\sigma\zeta}
{2(1+\sigma)(1+\zeta)}\pi^4Q.$$\par
\section{Envelope Equation}
In this section, we derive the equation which describes the nonlinear 
evolution of a slowly varying envelope of oscillatory travelling rolls
 near the critical Rayleigh number $R_0$. For values of $R$ close to $R_0$,
 $$R=R_0+\eps^2R^{(2)}\qquad (\eps\ll 1),$$
 we investigate the weakly nonlinear evolution of the wavepacket centered at
 the critical wavenumber $k_0$ and the corresponding frequency $\omega_0=
 \omega(k_0)$ by approximating $\bu,\bb,\Theta$ and $P$ as
 \begin{eqnarray}
 \bu&=&\eps\bu_1(\xi,\eta,\tau)E\left(\begin{array}
 {c}\cos(\pi z)\\ \cos(\pi z)\\ \sin(\pi z) \end{array}\right)+
 (\mbox{c.c.})+\eps^2\bu^{(2)}+\eps^3\bu^{(3)}+\cdots,\label{ex1}\\
 \bb&=&\eps\bb_1(\xi,\eta,\tau)E\left(\begin{array}{c}\sin(\pi z)\\ 
 \sin(\pi z)\\ \cos(\pi z) \end{array}\right)+(\mbox{c.c.})+
 \eps^2\bb^{(2)}+\eps^3\bb^{(3)}+\cdots,\label{ex2}\\
 \Theta&=&\eps\Theta_1(\xi,\eta,\tau)E\sin(\pi z)+(\mbox{c.c.})
 +\eps^2\Theta^{(2)}+\eps^3\Theta^{(3)}+\cdots,\label{ex3}\\
 P&=&P_1(\xi,\eta,\tau)E\cos(\pi z)+(\mbox{c.c.})+\eps^2P^{(2)}+
 \eps^3P^{(3)}+\cdots,\label{ex4}\\
E&\equiv& \exp[i(k_0x-\omega_0t)] ,\nonumber 
\end{eqnarray}
where (c.c.) denotes the complex conjugate of the previous term, and
\begin{equation}
\xi=\eps (x-\lambda t),\quad\eta=\eps y,\quad\tau=\eps^2t.\label{str}
\end{equation} 
The present ordering (\ref{str}) of the stretched variables $\xi,\eta,\tau$ 
is the same as one introduced in the derivation of the DS equation 
\cite{davey}. 
Substituting the expansions (\ref{ex1})-(\ref{ex4}) into Eqs(\ref{b1})-
(\ref{b4}) and using the method of multiple scales \cite{taniuti}, we obtain
 from the leading order equations(linearized equations)
\begin{eqnarray} 
\Theta_1&=&L_\theta w_1,\quad L_\theta=(\kappa^2-i\omega_0)^{-1},\nonumber\\
c_1&=&\pi L_cw_1,\quad L_c=(\zeta\kappa^2-i\omega_0)^{-1},\nonumber\\
P_1&=&\pi L_pw_1,\quad L_p=-\sigma(R_0L_\theta/\kappa^2+\zeta QL_c),
\nonumber\\
u_1&=&i\frac{\pi}{k_0}w_1,\quad a_1=-i\frac{\pi}{k_0}c_1,\nonumber\\
b_1&=&v_1=0, \nonumber
\end{eqnarray}
where $\kappa^2=k_0^2+\pi^2$ and the linear dispersion relation becomes

$$f\equiv \sigma\kappa^2-i\omega_0-\pi^2L_p-\sigma R_0L_\theta=0.$$

The second order field variables are expressed as follows
\begin{eqnarray}
w^{(2)}&=&w_1^{(2)}E\sin(\pi z),\nonumber\\
\Theta^{(2)}&=&\Theta_1^{(2)}E\sin(\pi z)+(\mbox{c.c.})+\Theta_0^{(2)}
\sin(2\pi z),\label{theta2}\\
c^{(2)}&=&[c_1^{(2)}\cos(\pi z)+\bar c_2^{(2)}E]E+(\mbox{c.c.}),\\
P^{(2)}&=&[P_1^{(2)}\cos(\pi z)+\{P_2^{(2)}\cos(2\pi z)+\bP_2^{(2)}\}E]E+
(\mbox{c.c.})\nonumber \\
& &+P_0^{(2)}\cos(2\pi z)+\bP_0^{(2)},\\
u^{(2)}&=&u_1^{(2)}E\cos(\pi z)+(\mbox{c.c.})+u_0^{(2)}\cos(2\pi z)+\baru_0^{(2)},
\\
a^{(2)}&=&a_1^{(2)}E\sin(\pi z)+(\mbox{c.c.})+a_0^{(2)}\sin(2\pi z),\\
v^{(2)}&=&v_1^{(2)}E\cos(\pi z)+\bar v_0^{(2)},\quad b^{(2)}=b_1^{(2)}E
\sin(\pi z),\label{vb2}
\end{eqnarray}
where $\baru_0^{(2)}$ and $\bar v_0^{(2)}$ are determined by
\begin{eqnarray}
(\dxi^2+\deta^2)\baru_0^{(2)}&=&\frac{\pi^2}{k_0^2\lambda}(1-\sigma\zeta Q
\pi^2|L_c|^2)\deta^2|w_1|^2,\label{u0}\\
(\dxi^2+\deta^2)\bar v_0^{(2)}&=&-\frac{\pi^2}{k_0^2\lambda}(1-\sigma\zeta Q
\pi^2|L_c|^2)\deta\dxi|w_1|^2.\nonumber 
\end{eqnarray}
The other second order amplitudes such as
 $\Theta_1^{(2)},\Theta_0^{(2)},c_1^{(2)} $ 
 and so on are given in Appendix and give 
 $$\lambda=[\partial_k \omega(k)]_{k=k_0}\equiv\partial_k \omega_0.$$
As shown in Appendix, the solvability condition for the third order 
variable $w_1^{(3)}$ yields the following equation of the envelope of the
first order vertical fluid velocity $w_1$ :
 \begin{equation}
 i\,\dtau w_1+\frac{\partial_k^2\omega_0}{2}\dxi^2w_1+\frac{\partial_k
 \omega_0}{2k_0}
 \deta^2w_1+\left(\frac{q}{\partial_{\omega_0}f}|w_1|^2-k_0
 \baru_0^{(2)}\right)w_1+\frac{\partial_{R_0}f}
 {\partial_{\omega_0}f}R^{(2)}w_1=0, \label{DS}
 \end{equation}
where $\partial_k\omega_0$ is real,  
while $\partial_k^2\omega_0\equiv(\partial_k^2\omega)_{k=k_0}$ has a 
complex value in general and 
 \begin{eqnarray}
 q&=&\sigma R_0\frac{k_0^2}{2\kappa^2}L_\theta L_\theta'+\sigma Q\frac
 {\zeta\pi^4}{2\zeta k_0^2-i\omega_0}\left[\frac{3k_0^2-\pi^2}{\kappa^2}
 |L_c|^2-Lc^2\right]
 \nonumber\\
 & &-\frac{8i\sigma Q\pi^6}{\kappa^2(Q+4\pi^2)}L_c''L_c
 \nonumber\\
 & &+\frac{iQ\pi^2}{Q+4\pi^2}\left(\pi^2\sigma\zeta QL_c^2+\sigma R_0\frac{k_0^2}
 {\kappa^2}L_\theta^2+\frac{k_0^2-3\pi^2}{\kappa^2}\right)L_c''.\label{q}
 \end{eqnarray}
A coupled system of equations (\ref{DS}) and (\ref{u0}) is the DS equation
 with complex coefficients, in which the real field $\baru_0^{(2)}$ 
  represents $z$ independent incompressive horizontal flows varying slowly.
  \\ \par
\section{Modulational Instability of Oscillatory Travelling Rolls}
For later conveniences, the DS equation with complex coefficients (a 
coupled system of equations (\ref{DS}) and (\ref{u0}) ) is rewritten as 
follows.
\begin{equation}
 i\,\pdt\Psi+\alpha\pdx^2\Psi+\beta\pdy^2\Psi+(\gamma|\Psi|^2+su)\Psi=i
 r\Psi, \label{DS1}
 \end{equation}
 \begin{equation}
 (\pdx^2+a\pdy^2)u= (b\pdx^2+c\pdy^2)|\Psi|^2,\label{DS2}
\end{equation}
where $\alpha,\beta$ and $\gamma$ are complex constants $(\alpha=\alpha'+
i\alpha'',\beta=\beta'+i\beta'',\gamma=\gamma'+i\gamma'')$, while $s,r,a,b$ 
and $c$ are real constants $(a=1,b=0,\beta''=0$ in the present
 case). A spatially uniform solution of Eqs.(\ref{DS1}) and (\ref{DS2}) is
 given by 
 $$\Psi=\Psi_0\equiv\psi_0\exp(-i\Omega t),\qquad u=0, $$
 where $|\psi_0|^2=r/\gamma''$ and $\Omega=-\gamma'|\psi_0|^2$. Setting
 \begin{eqnarray*} 
 \Psi&=&\Psi_0+\psi_1(t)\exp[i(\bok\cdot\br-\Omega_1t)]+
 \psi_2(t)\exp[i(-\bok\cdot\br-\Omega_2t)],\\
 u&=&u_1(t)\exp[i\{\bok\cdot\br-(\Omega_1-\Omega)t\}]\\
 & &+u_2(t)\exp[i\{-\bok\cdot\br-(\Omega_2-\Omega)t\}]+(\mbox{c.c.}),
 \end{eqnarray*}
 where $\bok=(k_x,k_y), \br=(x,y), 2\Omega=\Omega_1+\Omega_2$ and  
 linearizing Eqs.(\ref{DS1}) and (\ref{DS2}) with respect to $\psi_1$ and
  $\psi_2$, we have
 \begin{eqnarray}
 \frac{d\psi_1}{dt}&=&[r-i\{\alpha k_x^2+\beta k_y^2-\Omega_1-(\gamma+\hat
 \gamma)|\psi_0|^2\}]\psi_1
 +i\hat\gamma\psi_0^2\psi_2^*,\label{l1}\\
 \frac{d\psi_2^*}{dt}&=&[r+i\{\alpha^*k_x^2+\beta^*k_y^2-\Omega_2-(\gamma^*
 +\hat\gamma^*)|\psi_0|^2\}]\psi_2^*
 -i\hat\gamma^*{\psi_0^*}^2\psi_1,\label{l2}
 \end{eqnarray}
 where $\hat\gamma=\gamma+s(bk_x^2+ck_y^2)/(k_x^2+ak_y^2)$ and * denotes 
 the complex conjugate. The linear equations (\ref{l1}) and (\ref{l2}) have 
 exponentially growing solutions if the following condition is satisfied.
 \begin{equation}
 (\alpha'\gamma'+\alpha''\hat\gamma'')k_x^2+(\beta'\gamma'+\beta''
 \hat\gamma'')k_y^2>0.\label{inst}
 \end{equation}
 Since $a>0$ in the present case, Eq.(\ref{inst}) yields the following 
 instability criteria.
 \begin{equation}
 \hat\alpha\equiv\alpha'(\gamma'+sb)+\alpha''\gamma''>0,\label{alpha}
 \end{equation}
 or
 \begin{equation}
 \hat\beta\equiv\beta'(\gamma'+sc/a)+\beta''\gamma''>0,\label{beta}
 \end{equation}
 or
 \begin{equation}
 (\hat\alpha-\hat\beta/a)^2+\hat s[2(\hat\alpha+\hat\beta/a)+\hat s]>0,
 \quad\mbox{and}\quad \hat\alpha+\hat\beta/a+\hat s>0,\label{newcr}
 \end{equation}
 where $\hat s=s(c/a-b)(\alpha'-\beta'/a)>0$. The criterion (\ref{alpha}) or
 (\ref{beta}) is essentially the same as the case of the two-dimensional
  complex Ginzburg-Landau equation (Eq.(\ref{DS1}) with $s=0$). A new
   criterion (\ref{newcr}) comes from the coupling between convetive rolls 
   and horizontal imcompressible flows.If $a<0$, although this is not the 
   case in magnetoconvection, a spatially uniform solution of Eqs.
   (\ref{DS1}) and (\ref{DS2}) is shown to be always modulatinally unstable.
\section{Concluding Remarks}     
 In this paper, we have derived a envelope equation of oscillatory travelling
 rolls near a Hopf bifurcation, which is not a type of the NWS equation \cite
 {brand} but the DS equation with complex coefficients. The NWS type equation
 has a defect if the group velocity at the critical (bifurcation) point does
 not vanish. That is, it consists of different order terms: 
 the main term is linear and proportional to the group velocity, 
 while the other interesting terms such as nonlinear terms are of higher 
 order. Although our derivation is based on weakly nonlinear analysis with
 multiple scales similar to \cite{brand}, the present ordering of 
 stretched variables (\ref{str})  is different from the NW ordering 
  \cite{newell1} and yields the DS equation with complex coefficients 
 which consists of the same order terms.\\
 Analyzing the modulational instability  of a spatially independent 
 oscillatory solution of the DS equation with complex coefficients, we 
 obtain criteria of the modulational instability, which include not only 
 the known criterion for the complex Ginzburg-Landau equation ,which was
 first given in \cite{lange}, but also a new
 criterion due to the coupling between convective rolls and 
 horizontal incompressible flows.\\     
\par

\begin{flushleft}
{\Large \bf Appendix} 
\end{flushleft}
\appendix
\section{Higher Order Amplitudes}
In terms of the first order amplitudes and $w_1^{(2)}$, the second order 
amplitudes defined in Eqs.(\ref{theta2})-(\ref{vb2}) are given by
\begin{eqnarray*}
\Theta_1^{(2)}&=&L_\theta w_1^{(2)}-i\dot L_\theta\dxi w_1,\quad
c_1^{(2)}=\pi L_cw_1^{(2)}-i\pi\dot L_c\dxi w_1, \\
P_1^{(2)}&=&\pi L_pw_1^{(2)}-i\pi \dot L_p\dxi w_1,\quad
u_1^{(2)}=i\frac{\pi}{k_0}(w_1^{(2)}+\frac{i}{k_0}\dxi w_1),\\
a_1^{(2)}&=&-i\frac{\pi}{k_0}(c_1^{(2)}+\frac{i}{k_0}\dxi c_1),\quad
v_1^{(2)}=\frac{\pi^2}{k_0}\deta w_1,\quad
b_1^{(2)}=-\frac{\pi^2}{k_0}\deta c_1,\\ 
\dot L_\theta&=&\lambda\partial_{\omega_0} L_\theta+\partial_{k_0}L_\theta
,\quad\mbox{etc.},
\end{eqnarray*}
and 
\begin{eqnarray*}
\Theta_0^{(2)}&=&-\frac{L_\theta '}{2\pi}|w_1|^2,\quad
u_0^{(2)}=-2Q\frac{\pi^2}{k_0(Q+4\pi^2)}L_c''|w_1|^2,\quad a_0^{(2)}=\frac
{2\pi}{Q\zeta}u_0^{(2)},\\
P_0^{(2)}&=&(1+\frac{\sigma R_0L_\theta'}{4\pi^2})|w_1|^2+\frac{\sigma
\zeta Q\kappa^2}{2k_0^2}|c_1|^2,\\
\bP_0^{(2)}&=&\lambda \baru_0^{(2)}-\frac{\pi^2}{k_0^2}|w_1|^2-\frac
{\sigma\zeta Q}{2}(1-\frac{\pi^2}{k_0^2})|c_1|^2,\\
P_2^{(2)}&=&-\frac{\sigma\zeta Q\kappa^2}{4k_0^2}c_1^2,\quad
\bP_2^{(2)}=-\sigma\zeta Q[\frac{\pi}{2\zeta k_0^2-i\omega_0}w_1+\frac
{\kappa^2}{4k_0^2}c_1]c_1+\frac{\pi^2}{2k_0^2}w_1^2,\\
\bar c_2^{(2)}&=&\frac{\pi}{2\zeta k_0^2-i\omega_0}w_1c_1,\quad
Lc''=\mbox{Im}(L_c),\quad L_\theta'=\mbox{Re}(L_\theta), 
\end{eqnarray*}
Lengthy but straightforward calculations of the third order terms in Eqs.
(\ref{b1})-(\ref{b4}) give the following third order field variables
(envelopes) proportional to $E$ in terms of $w_1$ and $w_1^{(2)}$.
\begin{eqnarray*}
\Theta_1^{(3)}&=&L_\theta w_1^{(3)}-i\dot L_\theta\dxi w_1^{(2)}-(\ddot L_
\theta/2)\dxi^2w_1-L_\theta^2(\dtau-\deta^2)w_1\\
& &-ik_0L_\theta^2(\baru_0^{(2)}-u_0^{(2)}/2)w_1+\pi L_\theta 
\Theta_0^{(2)}w_1,\\
c_1^{(3)}&=&\pi[L_cw_1^{(3)}-i\dot L_c\dxi w_1^{(2)}-(\ddot L_c/2)
\dxi^2w_1-L_c^2(\dtau-\zeta\deta^2)w_1]\\
& &+ik_0L_c[a_0^{(2)}/2-\pi L_c(\baru_0^{(2)}
+u_0^{(2)}/2)]w_1-\pi L_c\bc_2^{(2)}w_{-1},\\
P_1^{(3)}&=&\pi[L_pw_1^{(3)}-i\dot L_p\dxi w_1^{(2)}-(\ddot L_p/2)\dxi^2w_1
-1/(2k_0)\partial_{k_0}L_p\deta^2w_1\\
& &+i\partial_{\omega_0}L_p\dtau w_1-(\sigma R^{(2)}/\kappa^2)L_\theta w_1]\\
& &+\sigma\zeta Q\pi[L_c-L_c^*(k_0^2-3\pi^2)/\kappa^2]\bar c_2^{(2)}w_{-1}\\
& &-[(i\sigma\zeta Q/2)\{(k_0^4+4k_0^2\pi
^2-\pi^4)/k\kappa^2\}L_ca_0^{(2)}+k_0\pi\partial_{\omega_0}L_p\bar u_0^{(2)}\\
& &-(ik_0\pi/2)\{\sigma\zeta QL_c^2-(\sigma R_0/\kappa^2)L_\theta^2-
(4/\kappa^2)\}u_0^{(2)}\\
& &+(\sigma R_0\pi^2/\kappa^2)L_\theta\Theta_0^{(2)}]w_1 ,
\end{eqnarray*}
and 
\begin{eqnarray}
fw_1^{(3)}&=&i\dot f\dxi w_1^{(2)}-i\partial_{\omega_0}f\dtau w_1
+(1/2)(\ddot f\dxi^2w_1+k_0^{-1}\partial_{k_0}f\deta^2w_1)\nonumber
\\& &+(k_0\partial_{\omega_0}f\baru_0^{(2)}-\partial_{R_0}f
R^{(2)})w_1-q|w_1|^2w_1,\label{A1}
\end{eqnarray}
where $q$ is given in Eq.(\ref{q}). Since
$$f=\dot f=0,\qquad\frac{\partial_{k_0}f}{\partial_{\omega_0}f}=-\partial_k
\omega_0,\qquad\frac{\ddot f}{\partial_{\omega_0}f}=-\partial_k^2\omega_0,$$
Eq.(\ref{A1}) gives Eq.(\ref{DS}).\\
\par
\begin{flushleft}
{\Large \bf Ackowledgement}
\end{flushleft}
One of authors (K.N.) wishes to thank Prof.N.Bekki,Nihon University,for his
introduction to magnetoconvection.\\
\par

\end{document}